\documentclass{article}
\usepackage[utf8]{inputenc}
\usepackage{amssymb,amsfonts,amsmath}
\usepackage{graphicx}
\usepackage{overpic}
\usepackage{rotating}
\usepackage{float}
\usepackage{color}
\usepackage{appendix}

\usepackage[margin=1in]{geometry}
\newcommand{\fref}[1]{{Fig.~\ref{fig:#1}}}
\newcommand{\Fref}[1]{{Figure~\ref{fig:#1}}}

\title{{\textbf{Geometric and Control Optimization of a \\Two Cross-Flow Turbine Array}}}
\author{Isabel Scherl$^{*}$, Benjamin Strom, \\  Steven L. Brunton, and Brian L. Polagye\\
\footnotesize{Department of Mechanical Engineering, University of Washington, Seattle, WA 98195, 
}}
\date{}
\begin{document}

\maketitle

%\blfootnote{$^*$ Corresponding author (ischerl@uw.edu).}
\begin{abstract}

Cross-flow turbines, also known as vertical-axis turbines, convert the kinetic energy in moving fluid to mechanical energy using blades that rotate about an axis perpendicular to the incoming flow. In this work, the performance of a two-turbine array in a recirculating water channel was experimentally optimized across sixty-four unique array configurations. For each configuration, turbine performance was optimized using \emph{tip-speed ratio control}, where the rotation rate for each turbine is optimized individually, and using \emph{coordinated control}, where the turbines are optimized to operate at synchronous rotation rates, but with a phase difference. For each configuration and control strategy, the consequences of co- and counter-rotation were also evaluated. We hypothesize how array configurations and control cases influence interactions between turbines and impact array performance.

\vspace{0.125in}

\end{abstract}

\section{Introduction}

Cross-flow turbines (i.e., vertical-axis turbines in wind) are an innovative technology for harnessing energy from wind and marine currents~\cite{salter2012coe,eriksson2008jrse} { that have recently seen a resurgence of interest~\cite{parker2016eif, dunne2015eif, bachant2015jot, bachant2016plos}.}
A principal advantage of cross-flow turbines is that dense arrays can outperform equivalent turbines in isolation~\cite{brownstein2016jrse, strom2018imej}. This complements other benefits of dense arrays, including increased power output per area~\cite{dabiri_2014}.  
Similar to flow control that enhances propulsion in biological systems (e.g., fish schooling ~\cite{fish2006ar, wu2011arfm, verma2018pnas}, flocking birds ~\cite{portugal2014nature, wu2011arfm}), augmented performance for cross-flow turbine arrays arises from beneficial interactions with the mean flow and coherent structures. 
{These mechanisms have also been explored for axial-flow turbines~\cite{shapiro2020arxiv}.} A common finding from prior investigations of cross-flow turbine arrays in field experiments~\cite{dabiri2011jrse,kinzel2012jot,kinzel_2015}, laboratory experiments~\cite{brownstein2016jrse, brownstein2019energies, ahmadi_2016}, and simulation~\cite{duraisamy_2014,zanforlin2016re,bremseth_2016,durrani_2011,chen_2017} is that power output increases when when the rotors are arranged in a side-by-side configuration. Chen et al.~\cite{chen_2017} considered the optimization of cross-flow turbine control in arrays computationally, but there has been no prior experimental work that considers simultaneous optimization of array configuration and control. 

In this work, we experimentally perform simultaneous optimization of turbine configuration and control with a two-rotor array with a constant free-stream velocity. 
We first consider a standard control method that independently optimizes the tip-speed ratios ($R \omega / U_{\infty}$) of each turbine, exploiting mean flow alterations to maximize array power output. Throughout, we refer to this control strategy as \textit{tip-speed ratio control}. We also introduce a new type of control with a synchronous rotation rate and a phase offset between the turbines in the array. We refer to this strategy, which coordinates the operation of turbines to exploit mean \textit{and} periodic flow field alterations, as \textit{coordinated control} . This strategy is motivated by the inherently periodic nature of cross-flow turbine fluid dynamics, in which the angle of attack encountered by the blades is continuously changing. 
The resulting unsteady aerodynamics can result in leading edge vortex separation~\cite{eldredge2019arfm} and dynamic stall~\cite{buchner2015jow, buchner2018jfm}.
Even at tip-speed ratios high enough to suppress dynamic stall~\cite{laneville1986josee, ferreira2009eif, buchner2018jfm, edwards2015we, fujisawa2001jow}, the angle of attack variation produces a phase-dependent pressure and velocity field in and around the rotor. As prior work has evaluated performance benefits for co-rotating and counter-rotating turbines~\cite{brownstein2019energies}, we consider both modalities in our experiments.  

\section{Methods}\label{methods}

\begin{figure}
\centering
\begin{overpic}[width=0.99\textwidth]{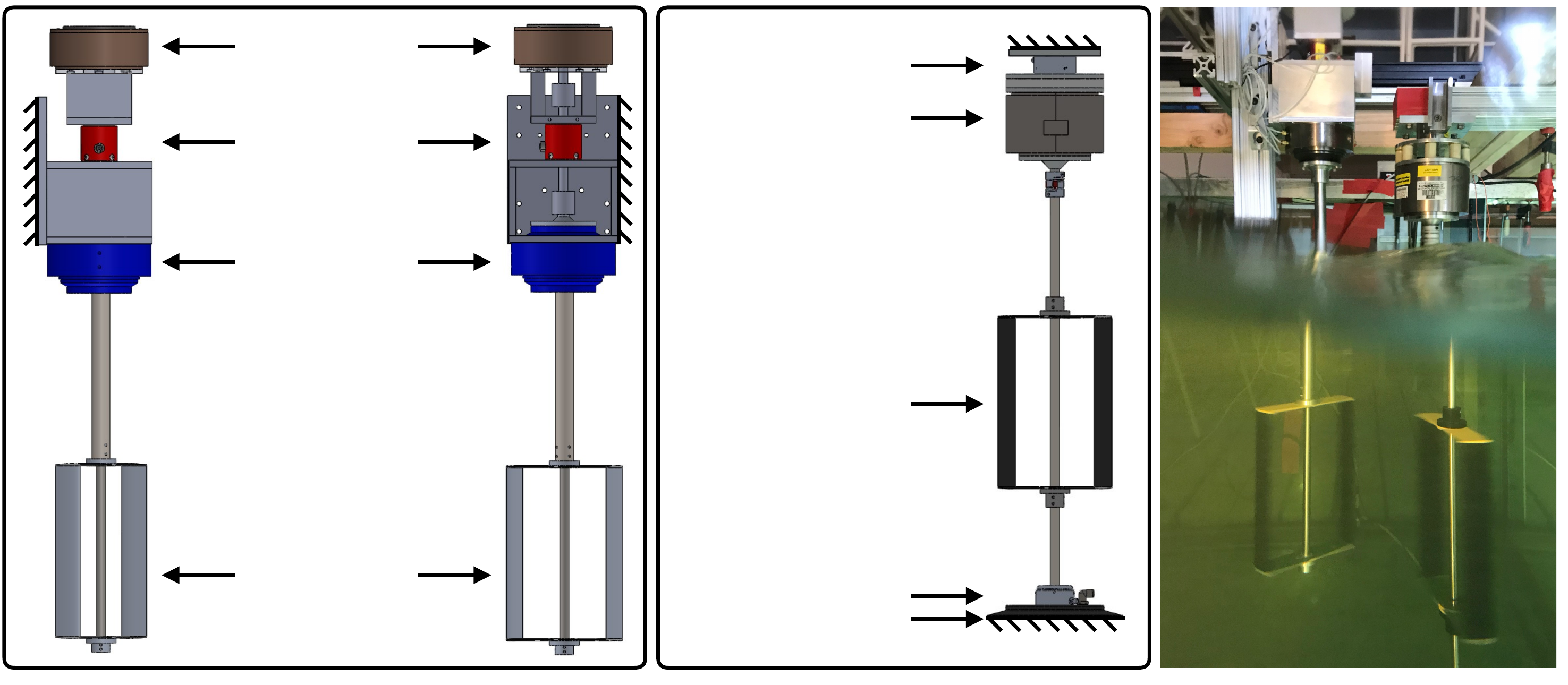}
\put(17.75,40.5){\textbf{motor}}
\put(17,35){\textbf{torque}}
\put(19,33){\textbf{cell}}
\put(19,27){\textbf{air}}
\put(17,25){\textbf{bearing}}
\put(17,7){\textbf{turbine}}
\put(18,5){\textbf{rotor}}

\put(48,39){\textbf{load cell}}
\put(50.25,35.75){\textbf{motor}}
\put(43,17.5){\textbf{turbine rotor}}
\put(48,5.25){\textbf{load cell}}
\put(43,3.5){\textbf{vacuum plate}}

\put(1.1,44.5){(a)}
\put(42.25,44.5){(b)}
\put(74.5,44.5){(c)}

\end{overpic}
\vspace{-.15in}
\caption{(a) Experimental set-up of the mobile, cantilevered turbine which consisted of a motor to enforce consistent tip-speed ratios and phase differences, a torque cell, an air bearing to absorb cantilever loads, and the rotor. (b) Experimental set-up of the stationary (or fixed) turbine, which consists of two load cells, a motor, the vacuum plate, and the rotor. (c) Photo of both turbines operating in the Bamfield Marine Science Centre flume.}\label{fig:exp}
\end{figure}
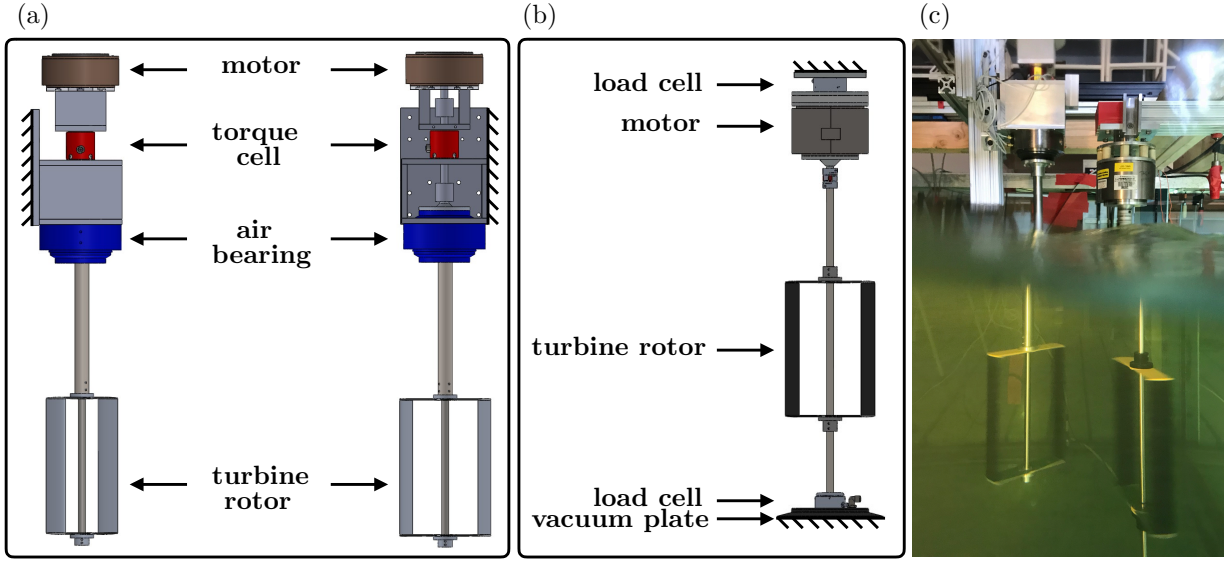

Array experiments were performed in the Bamfield Marine Science Centre recirculating water channel. During these tests, one turbine was fixed in space while the other turbine was cantilevered from a robotic gantry system (Velmex BiSlide) that allowed precise control of the rotor position in the streamwise and cross-stream directions. This enabled accurate and reproducible array geometries. 

The rotation rate of the stationary (or fixed) turbine (\fref{exp}b) was regulated by a servomotor ~\cite{polagye2019jrse} (Yaskawa SGMCS-05B with Yaskawa SGDV-05B3C41 drive). The motor had an internal encoder with over one million counts per revolution and was rigidly connected to a six-axis load cell (ATI Mini45) that measured all reaction forces and torques and was fixed to the flume structure. The turbine driveline was a 12.7 mm diameter stainless steel shaft that terminated, at the lower end, in a bearing that was fixed to a second six-axis load cell (ATI Nano25). This assembly was fixed to the bottom of the flume using a suction plate and scroll vacuum pump (Agilent IDP3) and has been used extensively in previous work~\cite{strom2017natenergy, strom2018jrse}. 
For the mobile turbine, elimination of the lower load cell and vacuum plate would have introduced significant cross-talk in an upper load cell due to the orders-of-magnitude difference between the thrust-induced moment and the relatively small rotary torque. As an alternative approach, as shown in \Fref{exp}a, the turbine driveline was supported by an air bearing with negligible friction (Professional Instruments Company Block-Head 4R low-inertia). For the mobile turbine, the rotation rate was regulated by a servomotor with internal encoder (Yaskawa SGMCS-02B3C41) and the reaction torque (equivalent to fluid torque) was measured by a torque cell (Futek F400). {A picture of both turbines operating in the Bamfield Marine Science Centre flume is shown in \Fref{exp}c}.

Both turbines had two blades with strut endplates to minimize parasitic losses~\cite{strom2018jrse}. Each turbine had a height of $H=0.23$ m, diameter of $D=0.172$ m, chord length of $c=0.04$ m, and symmetric NACA0018 blade profile. This turbine geometry has been used in past work~\cite{strom2017natenergy,strom2018jrse, polagye2019jrse,hunt2020jrse}. Due to the limited number of strut endplates, one turbine had struts with a NACA0008 profile, while the other used a NACA0016 profile. This gave rise to minor performance variations (\fref{paramspace}c). Both sets of struts had chord lengths that are equal to those of the blades (0.04 m). 

The Bamfield Marine Science Centre recirculating water channel was 10 m long, 2 m wide, and was filled to a dynamic depth of $h = 0.73 m$. The nominal flow speed was $U_{\infty}=0.6$ m/s with a turbulence intensity of approximately 2\%. The freestream flow velocity was measured upstream of the array using an acoustic Doppler velocimeter (Nortek Vector) at a rate of 64 Hz. {Despiking was performed using the method of Goring and Nikora~\cite{goring_2002}, with spikes replaced by interpolation.} The temperature was held at 18$\pm1$ $^\circ$C. These conditions corresponded to a chord-based Reynolds number of $Re_c = \frac{U_\infty c}{\nu}  \approx 22,000- 23,000$, where $\nu$ is the kinematic viscosity, and depth-based Froude number of $Fr = \frac{U_\infty}{\sqrt{gh}} \approx 0.22$, where $g$ is the gravitational acceleration or $9.81m/s^2$. The former placed the turbines in a transitional regime where performance varied with local $Re_c$~\cite{miller2018jfm}. The blockage ratio, defined as the projected area of the turbines normalized by the area of the channel, ranged from 2.8\% (turbines directly in-line) to 5.6\% (turbines side-by-side).  

The test matrix of relative rotor positions sampled is given in \Fref{paramspace}a. Arranged on a polar grid, positions ranged from side-by-side ($x = 0$) to $x = 3.61D$ in the streamwise direction, and a maximum spacing of $1.83D$ in the cross-stream direction. The minimum distance between rotor centers was $1.1D$. As discussed in Sec. \ref{expdesign}, the turbine interactions were not symmetric about $Y/D = 0$.

\begin{figure}
\centering
\begin{overpic}[width=0.99\textwidth]{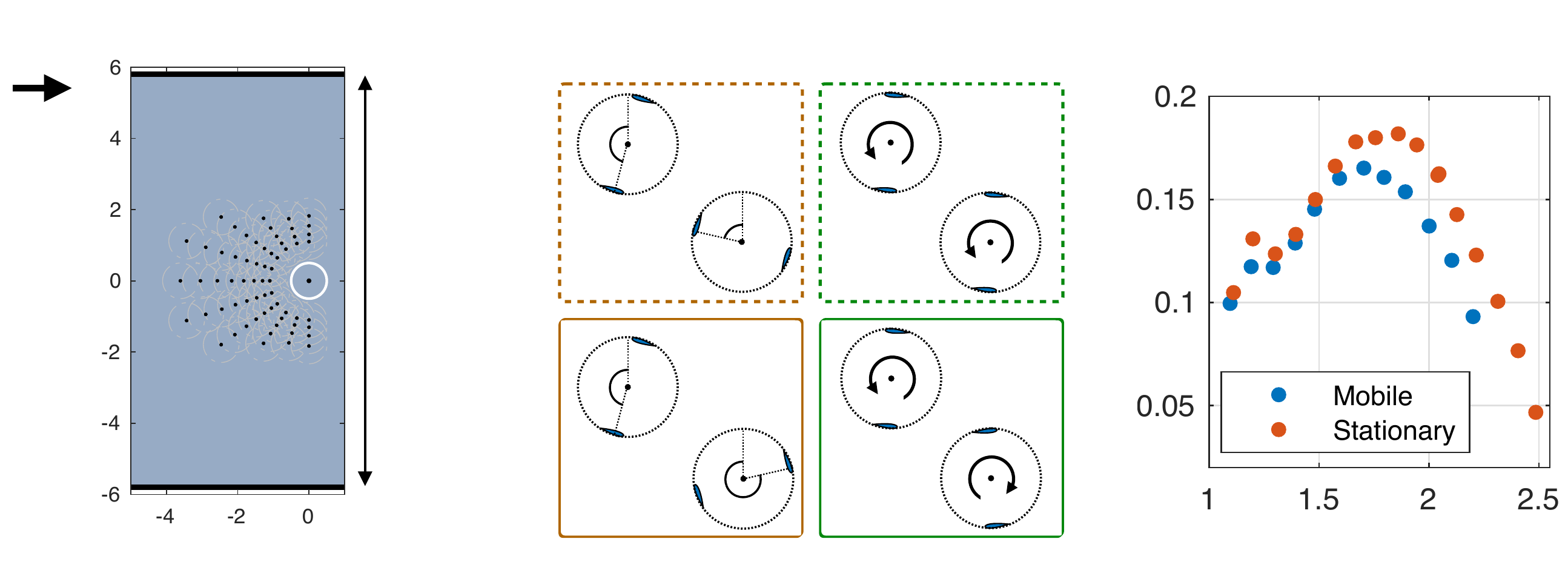}
\put(-.2,33){$U_{\infty}$}

\put(37.5,11){$\scriptstyle\theta_1$}
\put(45,5){$\scriptstyle \theta_2$}

\put(37.5,26.5){$\scriptstyle \theta_1$}
\put(45,23){$\scriptstyle\theta_2$}

\put(56.5,10.5){$\scriptstyle\lambda_1$}
\put(62.75,4.25){$\scriptstyle \lambda_2$}

\put(56.5,25.5){$\scriptstyle \lambda_1$}
\put(62.75,19.5){$\scriptstyle\lambda_2$}

\put(36,19){$\scriptstyle\phi = \theta_1 - \theta_2$}
\put(36,21){$\scriptstyle\lambda = \lambda_1 = \lambda_2$}
\put(45.5,29){$[\phi, \lambda]$}

\put(36,5.5){$\scriptstyle\lambda = \lambda_1 = \lambda_2$}
\put(36,3.5){$\scriptstyle\phi = -\theta_1 - \theta_2$}
\put(45.5,14){$[\phi, \lambda]$}

\put(33,6){\begin{sideways}Counter- \end{sideways}}
\put(33,23){\begin{sideways}Co- \end{sideways}}

\put(38,32.5){Coordinated}
\put(52.5,32.5){Tip-speed ratio}

\put(13, 0.5){{X/D}}
\put(3.2, 16.8){\begin{sideways}{Y/D}\end{sideways}}

\put(24.5,20){\textbf{Flume}}
\put(24.5,18){\textbf{width:}}
\put(25,15.6){11.6D}

\put(88, 1){$\lambda$}
\put(70.5, 18){\begin{sideways}$C_{P}$\end{sideways}}

\put(61,14){$[\lambda_1, \lambda_2]$}
\put(61,29){$[\lambda_1, \lambda_2]$}

\put(7,35){(a)}
\put(34,35){(b)}
\put(72,35){(c)}

\end{overpic}
\vspace{-.1in}
\caption{(a) Experimental matrix showing the  position where each control strategy was tested. (b) Each array  configuration tested, where one turbine was fixed at $X/D = Y/D=0$ and the other turbine was tested at each prescribed polar grid location. (c) Performance curve for each turbine operating in isolation.}\label{fig:paramspace}
\end{figure}

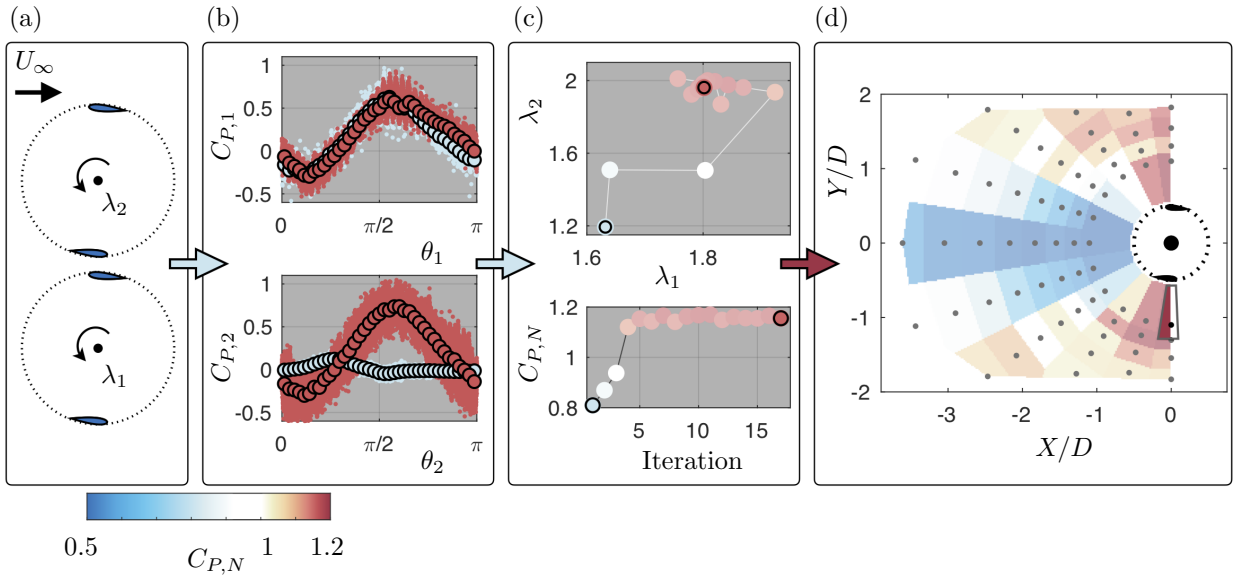
\begin{figure}
% \vspace{0.2in}
\centering
\begin{overpic}[width=0.99\textwidth]{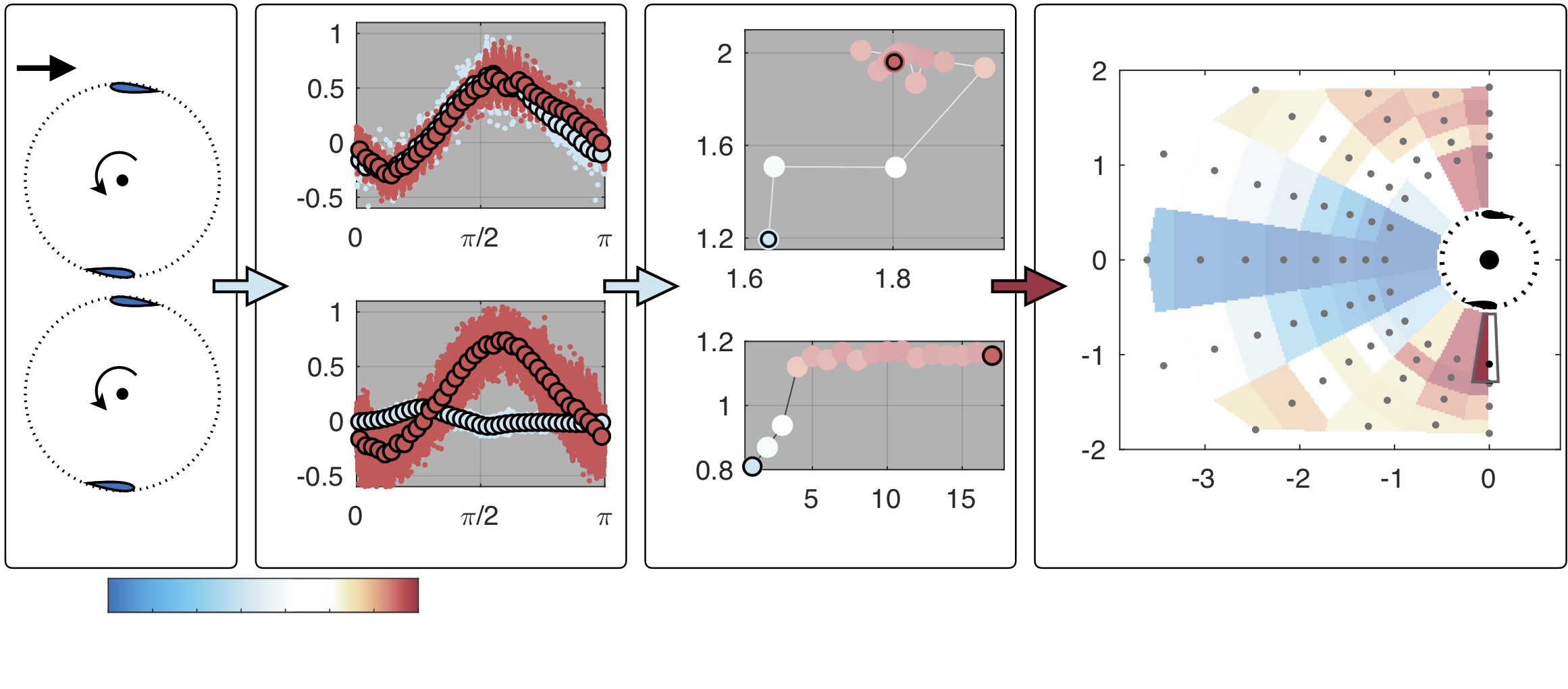}
\put(0.5, 45){(a)}
\put(16.5, 45){(b)}
\put(41.5, 45){(c)}
\put(66, 45){(d)}
\put(25,2){1.2}
\put(1,41.5){$U_{\infty}$}
\put(21,2){1}
\put(5,2){0.5}
\put(15,1){$C_{P,N}$}
\put(34,26){$\theta_{1}$}
\put(34,9){$\theta_{2}$}
\put(17,34){\begin{sideways}$C_{P,1}$\end{sideways}}
\put(17,16){\begin{sideways}$C_{P,2}$\end{sideways}}
\put(53,24){$\lambda_{1}$}
\put(42,37){\begin{sideways}$\lambda_{2}$\end{sideways}}
\put(42,16){\begin{sideways}$C_{P,N}$\end{sideways}}
\put(8,16){$\lambda_{1}$}
\put(8,30){$\lambda_{2}$}
\put(52,9){Iteration}
\put(84,10){$X/D$}
\put(67,31){\begin{sideways}$Y/D$\end{sideways}}
\end{overpic}
\vspace{-.15in}
\caption{Optimization for one geometric configuration of co-rotating turbines under tip-speed ratio control contrasting initial case (light blue) with optimized case (red). From left to right:  (a) the co-rotating turbines and direction of free-stream flow; (b) performance comparison of initial and optimized cases for each turbine as a function of the angular position of each turbine ($\theta= 0$ when the blade is pointing directly upstream); (c) top shows the optimization path and bottom shows the performance evolution during optimization; (d) resulting performance of selected array geometry on a heat map corresponding to all test positions. }\label{fig:flowInfo}
\end{figure}

\subsection{Performance Metrics}

The coefficient of performance is the ratio of the power produced by the turbine to the kinetic power in the free-stream passing through the turbine's projected area and expressed as %
\begin{equation}
C_p = \frac{P}{\frac{1}{2}\rho U_{\infty}^3HD} = \frac{\omega \tau}{\frac{1}{2}\rho U_{\infty}^3HD},
\end{equation}
where $P$ is turbine's mechanical power, $\rho$ is the fluid density, $U_\infty$ is the freestream flow velocity, $H$ is the turbine height, $D$ is the turbine diameter, $\omega$ is the turbine rotation rate, and $\tau$ is the turbine torque. This metric is extended to evaluate non-dimensional array performance.
In experiments, the two turbine rotors were tested separately to establish a baseline performance for comparison. Their individual peak performances are denoted by $C_{P,1}^*$ and $C_{P,2}^*$, where an asterisk denotes performance at the optimized tip-speed ratio (peak of the $C_P-\lambda$ curves in \Fref{paramspace}c).  The array performance is evaluated relative to the sum of the peak power that the two turbines would produce in isolation. 
The normalized array performance is then
\begin{equation}\label{eq:cc_perf}
C_{P,N} = \frac{C_{P,1}+C_{P,2}}{C_{P,1}^*+C_{P,2}^*}.
\end{equation}
{If mean inflow was perfectly constant across all tests, this would be equivalent to the power produced by the turbines in the array divided by the power produced by the turbines in isolation. Our choice of a non-dimensional performance metric controls for small fluctuations in the free stream velocity.} 

\subsection{Turbine Control}

At each of the locations illustrated in \Fref{paramspace}a, two control strategies are optimized to maximize array power output for both co- and counter-rotating turbines. 

\textit{Tip-speed ratio control} %{\color{red} (referred to as ``TSR'' in figures)} 
is characterized by the tip-speed ratio or non-dimensional rotation rate, given by
\begin{equation}
\lambda = \frac{\omega R}{U_{\infty}},
\end{equation}
where $R$ is the turbine radius.
For tip-speed ratio control, the rotation rates of each of the rotors are optimized simultaneously, assuming a nominally constant free-stream velocity. 
The tip-speed ratio for both turbines $(\lambda_1, \lambda_2)$ is defined using the \textit{freestream} flow velocity ($U_{\infty}$) which may not match the flow incident to the turbine, particularly when the rotors are in-line. 
The objective of this control scheme is to optimize rotor operation in the mean flow field induced by the rotors. This optimization problem is non-trivial, as there is a co-dependence between turbine rotation rates, the resulting flow field, and the array power output. 

In \textit{coordinated control}%{\color{red}(referred to as ``CC'' in figures)}
, the angular velocities, or tip-speed ratios, of the two turbines are locked to the same value ($\lambda = \lambda_1= \lambda_2$). For co-rotating turbines, the angular blade offset, or phase difference, is 
\begin{equation}
\phi = \theta_1 - \theta_2
\end{equation}
where $\theta_1$ and $\theta_2$ are the angular positions of mobile and stationary turbines respectively and $\theta = 0$ when one blade is pointing directly upstream. For counter-rotating turbines, the phase difference is defined as 
\begin{equation}
\phi = -\theta_1 - \theta_2.
\end{equation}
This results in another two-parameter optimization, this time of $\lambda$ and $\phi$. A closed-loop controller is used to maintain a constant $\phi$ while testing a specific pair of parameter values.

Optimization of the control parameters is performed using the Nelder-Mead algorithm~\cite{nelder_1965}. 
To improve the convergence rate of the optimization, initial simplex values are chosen, in part, based on data already collected. The first simplex point is the optimum control set point found for any array configuration so far. The second point is the optimum control set point of an array configuration with an upstream turbine position within $1D$ of the current mobile position. If that control set point is not suitably different from the first simplex point a psuedo-random second simplex vertex is chosen, ensuring sufficient distance in the parameter space. The third simplex point is always psuedo-random, but with a lower bound set so that the initial simplex volume is sufficiently large. Optimization is halted when the simplex reached a minimum volume or if 30 control set points have been tested since the last improvement in array performance. 
{\Fref{flowInfo} shows how the optimization was performed for each case. For a given geometry, control strategy, and rotation scheme (\fref{flowInfo}a), the array is tested for 25 seconds. Performance is computed (\fref{flowInfo}b) and the Nelder-Mead optimizer is used to determine the next point. This is repeated until convergence (\fref{flowInfo}c).  The performance resulting from the optimized control parameters is shown as a heat map for each array geometry in \Fref{flowInfo}d}.

\section{Results and Discussion}

\subsection{Co-rotation} \label{coarrays}

{Array performance for co-rotating turbines under tip-speed ratio control and coordinated control} is shown in \Fref{co_eff}. {Both turbines are rotating counterclockwise in the co-rotating arrays.} Normalized array performance values greater than 1 (red) represent configurations where the array is outperforming {the two turbines in isolation}. For both cases, when turbines are in similar $Y/D$ locations, performance suffers, but when they are in dissimilar $Y/D$ locations, performance is similar to or greater than isolated turbines. The co-rotating tip-speed ratio control case had the highest array performance of any array geometry, control type, and {rotation direction, with turbines situated side-by-side with a $1.1D$ center-to-center spacing achieving a non-dimensional array performance of 1.3 (i.e., a 30 \% increase in power output relative to turbines in isolation).}

\begin{figure}
\vspace{-.15in}
\centering
\begin{overpic}[width=\textwidth]{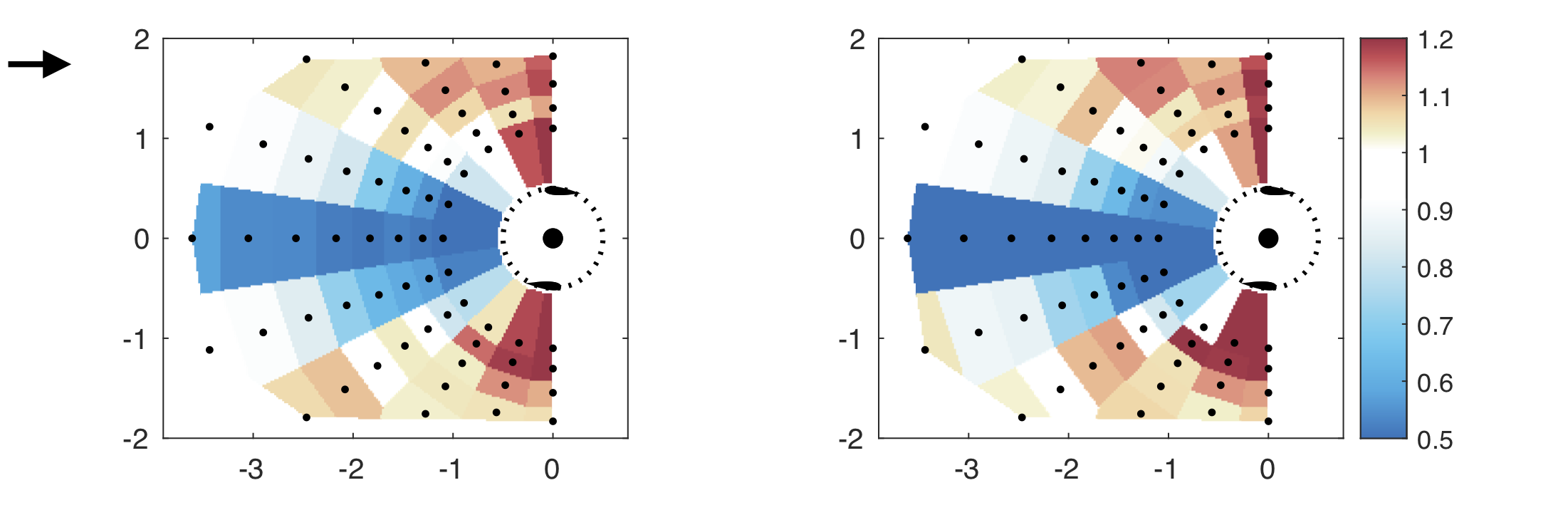}
\put(5,14.75){\begin{sideways}{$Y/D$}\end{sideways}}
\put(23.25, -1){{$X/D$}}
\put(69.25, -1){{$X/D$}}
\put(94.5, 16){$ C_{P,N}$}

\put(4.9, 29.5){(a)}
\put(50, 29.5){(b)}

\put(-0.5, 30){$U_\infty$}

\end{overpic}
\caption{Array performance for co-rotating arrays under (a) tip-speed ratio control and (b) coordinated control. The fixed turbine is located at $Y/D = X/D = 0$ and the black dot denotes the location of the upstream turbine. Array performance greater than 1 (red) indicates that the array is outperforming the turbines in isolation. The color map contains a dead-band surrounding unity to highlight variations.}% due to uncertainty and noise in measurements.}
\label{fig:co_eff}
\end{figure}

\begin{figure}[t]
\vspace{-.15in}
\centering
\begin{overpic}[width=\textwidth]{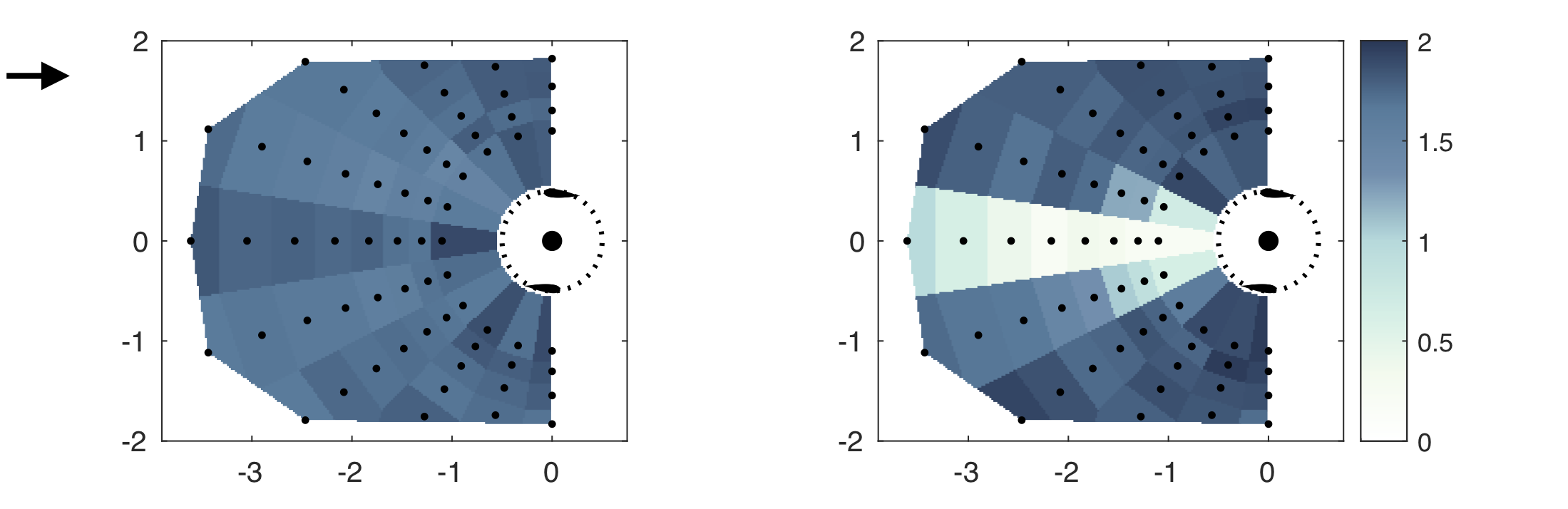}
\put(4.75,14.75){\begin{sideways}{$Y/D$}\end{sideways}}
\put(23.25, -1){{$X/D$}}
\put(69.25, -1){{$X/D$}}

\put(4.75, 29.5){(a)}
\put(50, 29.5){(b)}
\put(-0.5, 29.5){$U_\infty$}

\put(94.5, 16.25){$ \lambda$}
\end{overpic}
\vspace{-0.15in}
\caption{Optimized parameters for co-rotating turbines under tip-speed ratio control. (a) Upstream turbine tip-speed ratio for a given array layout and (b) downstream turbine tip-speed ratio.}\label{fig:co_tsr}
\end{figure}

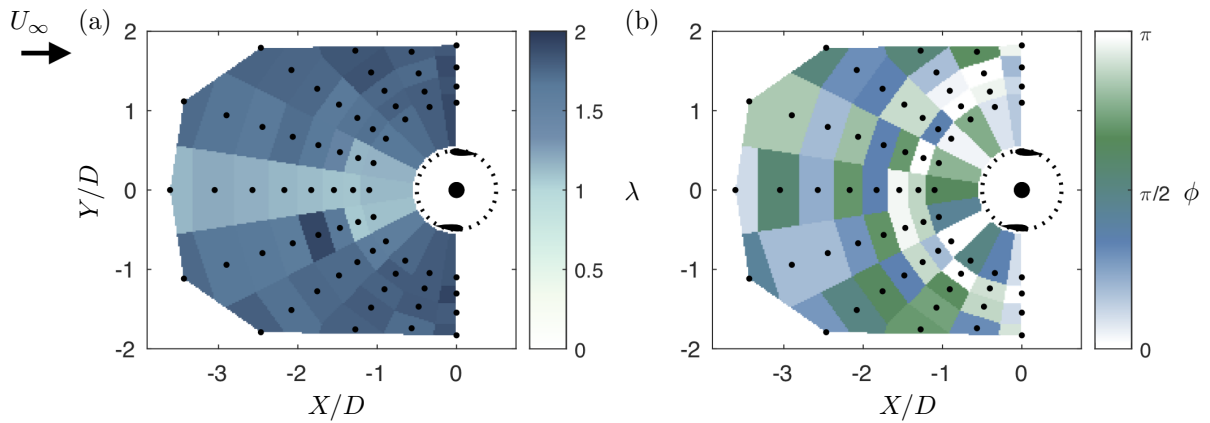
\begin{figure}
\centering
\begin{overpic}[width=\textwidth]{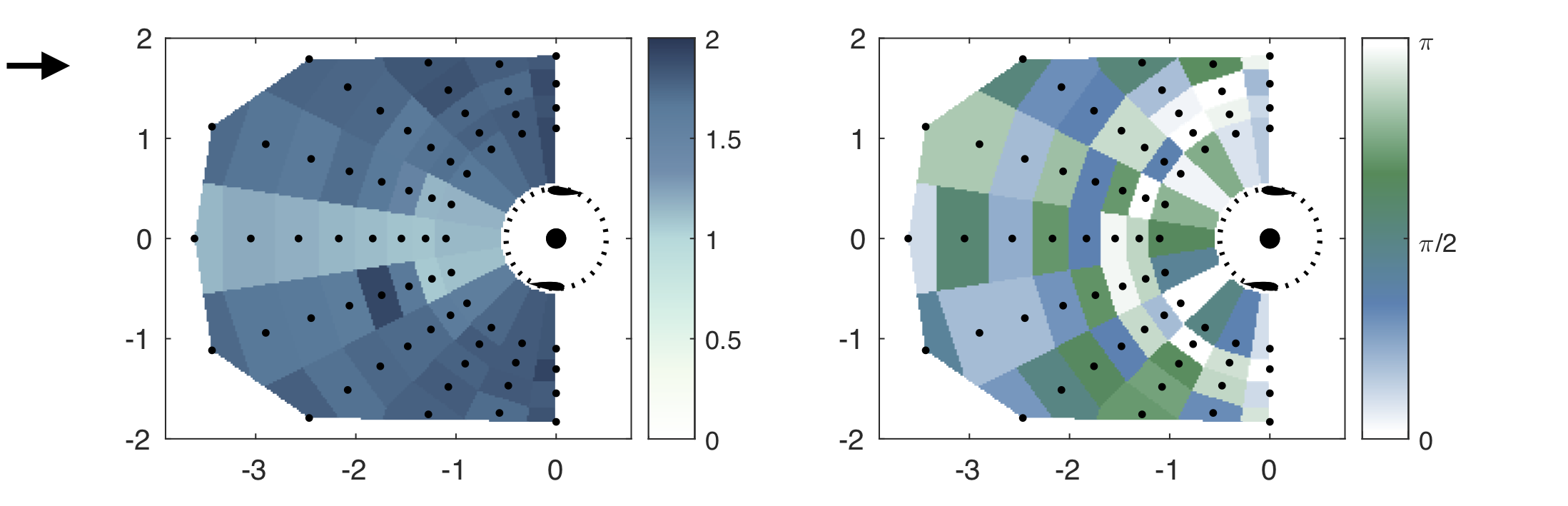}
\put(5,14.75){\begin{sideways}{$Y/D$}\end{sideways}}
\put(23.5, -1){{$X/D$}}
\put(69.5, -1){{$X/D$}}
\put(49, 16.25){$ \lambda$}
\put(94, 16.25){$ \phi$}
\put(5, 30){(a)}
\put(49, 30){(b)}
\put(-0.5, 30){$U_\infty$}
\vspace{-0.15in}
\end{overpic}
\caption{Optimized parameters for co-rotating turbines under coordinated control:  (a) optimized tip-speed ratio for the turbine pair and (b) optimized phase difference.}\label{fig:co_cc}
\end{figure}

{To better understand the drivers for array performance, we first consider the} optimized tip-speed ratios for the upstream and downstream turbines (\Fref{co_tsr}). For all configurations, the upstream turbine optimization converges to a relative narrow range of $\lambda$ ($1.5 \leq \lambda \leq 2$).  Since the optimal $\lambda$ in isolation is $1.7-1.9$ (\fref{paramspace}), this suggests that the upstream turbine is relatively unaffected by the presence of the downstream turbine. The downstream turbine optimization converges to a similar $\lambda$ when the turbines are at dissimilar $Y/D$ locations. When the downstream turbine is located at $Y/D = 0$, its optimized tip-speed ratio trends toward zero as $X/D \rightarrow 0$. This is likely a consequence of the reduction in local inflow velocity when the downstream turbine is located directly in the wake of the upstream turbine {and in such close proximity} relatively little wake recovery can occur~\cite{araya2017jfm}. In addition, this reduction in inflow velocity reduces the local Reynolds number, reducing $C_P$ at a given $\lambda$~\cite{miller2018jfm}.

The optimized parameters for coordinated control are shown in \Fref{co_cc}. The optimal values of $\lambda$ exhibit a qualitatively similar pattern to the downstream turbine under tip-speed ratio control (i.e., primarily affected when the turbines are at similar $Y/D$ locations). 
There is, however, evidence of radial striations in the optimal phase. Since, at these tip-speed ratios, the upstream turbine is periodically shedding vortices~\cite{snortland2019ewtec}, these radial striations could be indicative of meaningful interaction (interception or avoidance) with the coherent structures from the upstream turbine.

%For the coordinated control parameters (shown in Appendix \ref{app}, \fref{co_cc}), the optimized tip-speed ratio for the array has a qualitatively similar pattern to that of the downstream turbine in tip-speed ratio control {\color{blue}(\fref{co_tsr}b)}. However, the optimized tip-speed ratio values are greater {\color{blue}than for tip-speed ratio control} when the turbines are in similar $Y/D$ positions. %This trend matches the results for the counter-rotating tip-speed ratio control case that is explored in the next section. 

\subsection{Counter-rotation}\label{counterarrays}

\begin{figure} [t]
\vspace{-0.15in}
\centering
\begin{overpic}[width=\textwidth]{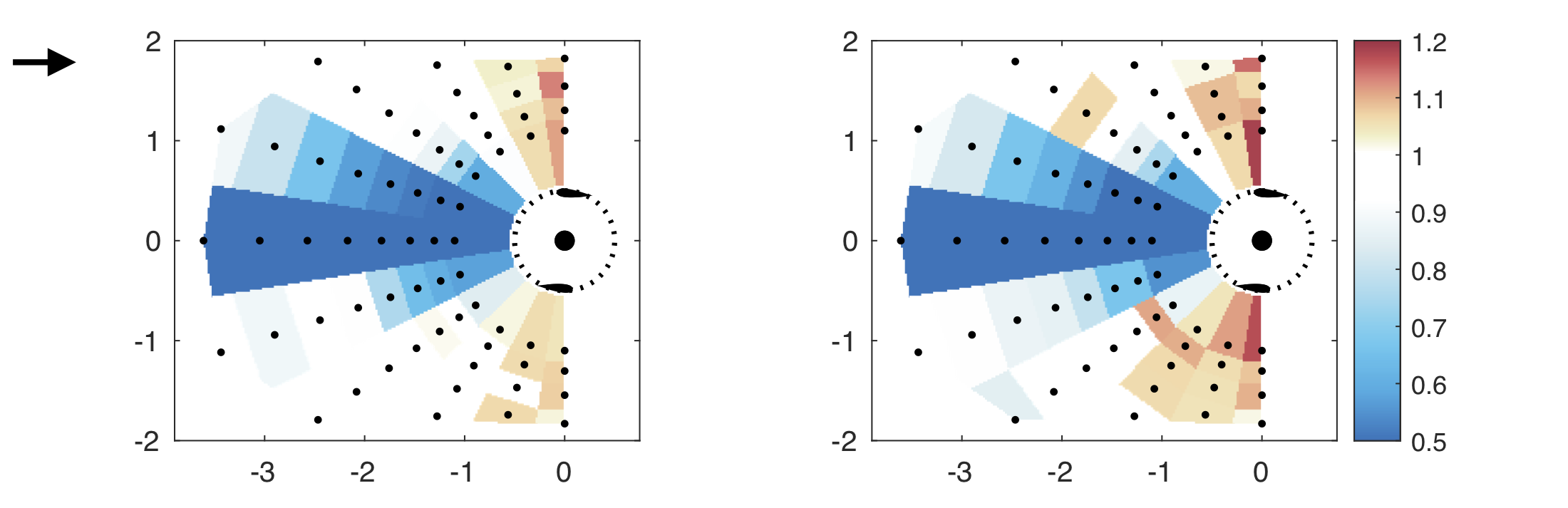}
\put(5,14.75){\begin{sideways}{$Y/D$}\end{sideways}}
\put(24, -1){{$X/D$}}
\put(68.75, -1){{$X/D$}}
\put(94, 16){$C_{P,N}$}
\put(5.25, 29.5){(a)}
\put(50, 29.5){(b)}
\put(-0.5, 30){$U_\infty$}
\end{overpic}

\caption{Array performance for counter-rotating arrays under (a) tip-speed ratio control and (b)coordinated control.}\label{fig:counter_eff}
\end{figure}
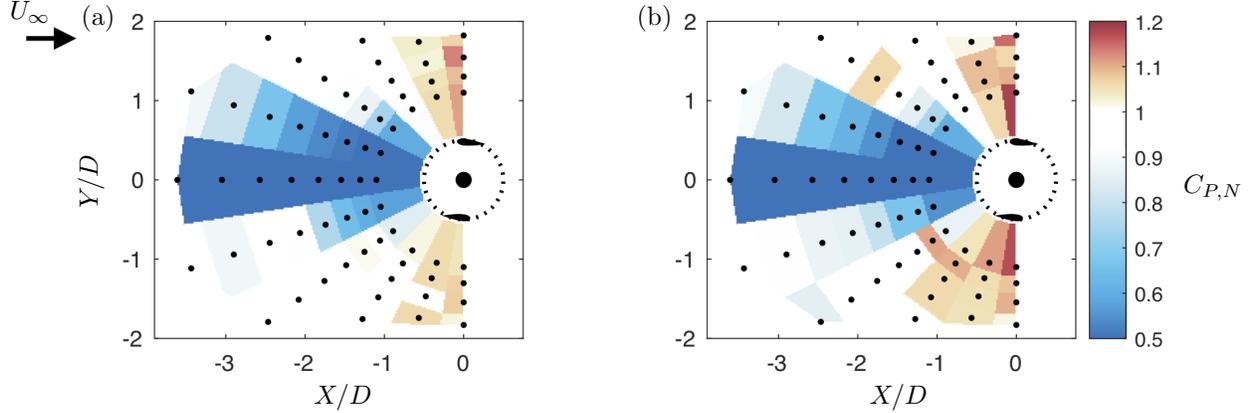

For the counter-rotating arrays, the upstream (mobile) turbine is rotating clockwise while the downstream (stationary) turbine remains rotating counterclockwise. Array performance trends for theses arrays, shown in \Fref{counter_eff}, are similar to the co-rotating arrays with array performance improving when the turbines are at similar $X/D$ positions and suffering when they are at similar $Y/D$ values. 

The counter-rotating tip-speed ratio control optimization values (Appendix \ref{app}, \fref{counter_tsr}) are also similar to the co-rotating values (\fref{co_tsr}). Specifically, for the upstream turbine, the optimized $\lambda$ are all similar to the isolated optima. When the turbines are located in the same $Y/D$ position, the downstream turbine is optimized to operate at a lower $\lambda$. The coordinated control values (shown in Appendix \ref{app}, \fref{counter_cc}) have a qualitatively similar pattern to the co-rotating coordinated control values (\fref{co_cc}). The tip-speed ratios are $1.7-1.9$ when the turbines are in dissimilar $Y/D$ locations, trend lower in similar $Y/D$ locations, and the phase difference shows evidence of radial striations. 

\subsection{Counter- vs. co-rotating arrays}

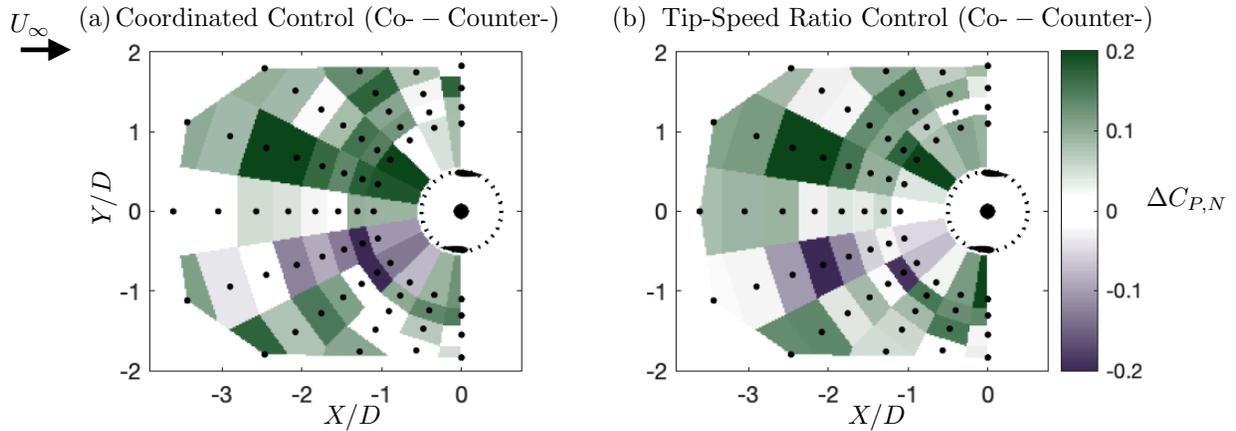
\begin{figure}
\centering
\begin{overpic}[width=\textwidth]{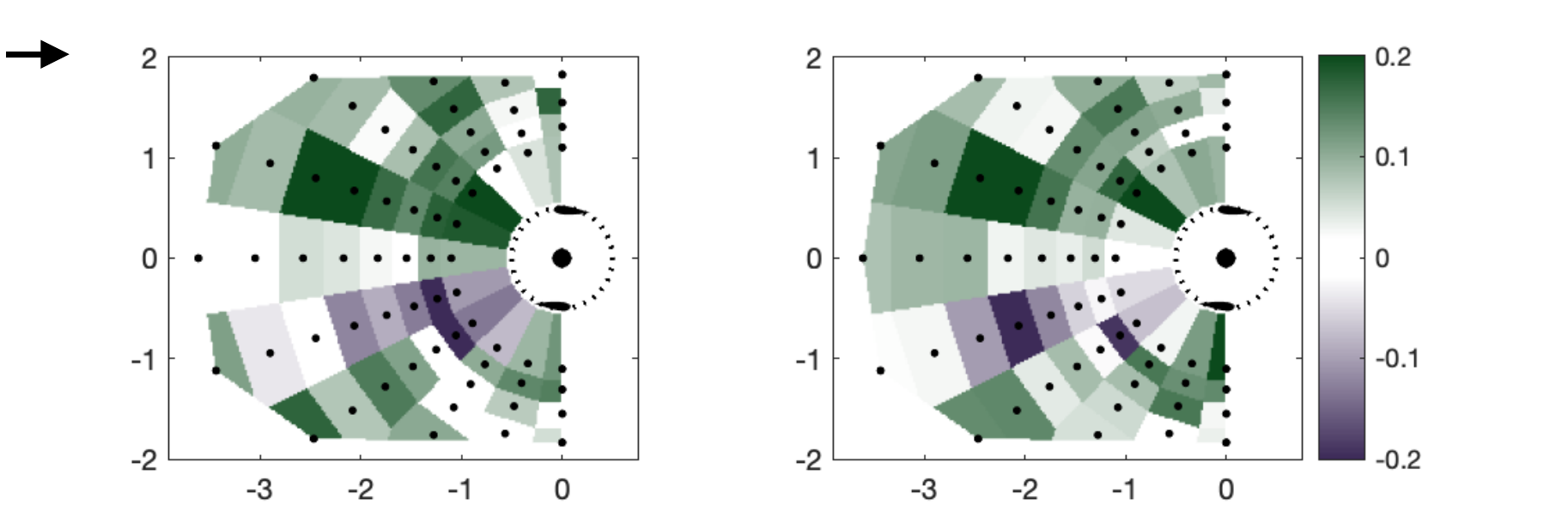}
\put(6,15){\begin{sideways}{$Y/D$}\end{sideways}}
\put(24.75, -1){{$X/D$}}
\put(67, -1){{$X/D$}}
\put(8, 31){Coordinated Control $(\text{Co-} -\text{Counter-})$}
\put(52, 31){Tip-Speed Ratio Control $(\text{Co-} -\text{Counter-})$}
\put(91, 16.5){$\Delta C_{P,N}$}
\put(5, 31){(a)}
\put(48, 31){(b)}
\put(-0.5, 30.5){$U_\infty$}

\end{overpic}
\caption{Array efficiency comparison ($\boldsymbol \Delta C_{P,N} = C_{P,N}^{\text{Co-}} - C_{P,N}^{\text{Counter-}}$) of co-rotating versus counter rotating arrays for (a) coordinated control and (b) tip-speed ratio control. Green denotes a co-rotating array out-performing its counter-rotating counterpart.}\label{fig:rotation_comp}
\end{figure}

As shown in \Fref{rotation_comp}, co-rotating arrays generally outperform counter-rotating arrays though the underlying mechanisms causing this trend are not immediately apparent. There is an exception where counter-rotating arrays outperform co-rotating arrays along a single vector. This can be explained by the turbine wake structure. In the $X-Y$ plane, cross-flow turbine wakes have an {asymmetric, skewed velocity deficit}, as shown in \Fref{rot}a~\cite{strom2019thesis}. For all cases tested, the stationary turbine (at $X/D = Y/D = 0$) is rotating counterclockwise, while the upstream (mobile) turbine is rotating counterclockwise in a co-rotating array and clockwise in a counter-rotating array. 

%When the turbines are co-rotating (i.e. the mobile and fixed turbines are both rotating counterclockwise), the wake from the upstream (mobile) turbine skews towards the positive $Y/D$ direction, as shown in \Fref{rot}b. For the locations where the co-rotating case performs worse than the counter-rotating case (purple regions in \Fref{rotation_comp}), the downstream turbine is likely situated in the region with the maximum velocity deficit. Conversely, when the upstream turbine is rotating clockwise (counter-rotating case), the wake skews towards the negative $Y/D$ direction (\fref{rot}c), such that the downstream turbine is in the wake of the upstream turbine for the cases that the co-rotating turbines perform best (dark green regions in \Fref{rotation_comp}).

{For the co- and counter-rotating cases, the upstream turbine rotates counterclockwise and clockwise, respectively. Therefore, as illustrated in \Fref{rot}, the upstream turbine wake deflection is in the $+Y/D$ direction for co-rotation and in the $-Y/D$ direction for counter-rotation. As a result, co-rotation underperforms counter-rotation in a band of $-Y/D$ upstream turbine positions ({purple} in  \Fref{rotation_comp}) while counter-rotation underperforms co-rotation for the mirrored $+Y/D$ upstream turbine positions ({dark green} in  \Fref{rotation_comp}). {For both cases, this is consistent with alternating directions of wake velocity skew for counterclockwise and clockwise rotation of the upstream turbine. As turbine separation increases beyond $X/D=-2$, this effect is reduced by wake mixing.}}

%{\color{red} The cause of the trend of co-rotating arrays generally outperforming counter-rotating arrays is not fully clear and would benefit from future exploration. in most cases, these differences are less than $0.1 \Delta C_{P,N}$.}

\begin{figure}
\centering
%\vspace{0.1in}
\begin{overpic}[width=\textwidth]{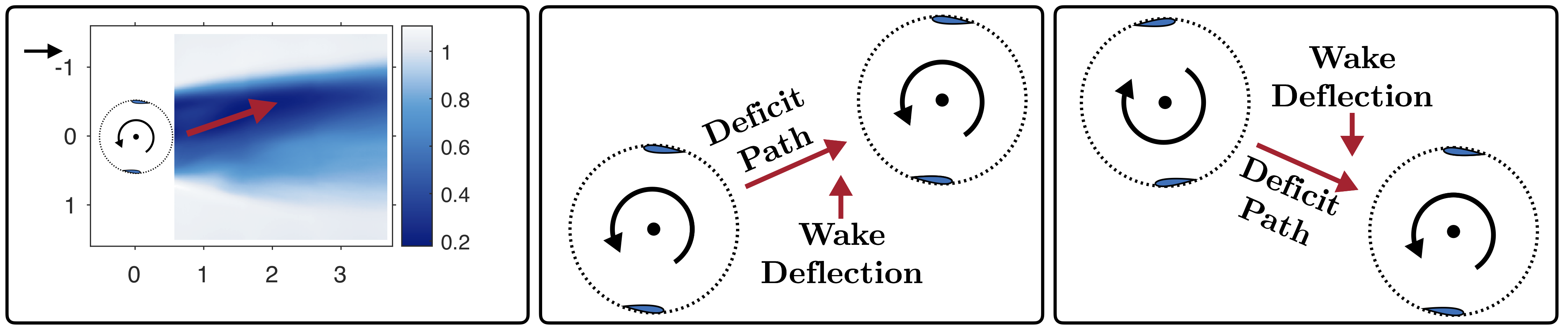}

\put(13.5, 1.45){{$\scriptstyle X/D$}}
\put(1.25, 10.5){\begin{sideways}{{$ \scriptstyle Y/D$}}\end{sideways}}

\put(30.35,17.5){$\frac{\text{U}}{\text{U}_\infty}$}

\put(1,22){(a)}
\put(35,22){(b)}
\put(68,22){(c)}
\put(0.75, 18.75){$\scriptstyle U_\infty$}

\end{overpic}
\caption{(a) Mean wake velocity of a cross-flow turbine rotating counterclockwise with the {skewed velocity deficit} denoted with a red arrow~\cite{strom2019thesis}, (b) deficit path and wake deflection direction for the co-rotating turbine array (c) deficit path and wake deflection direction for the counter-rotating turbine array.}\label{fig:rot}
\end{figure}
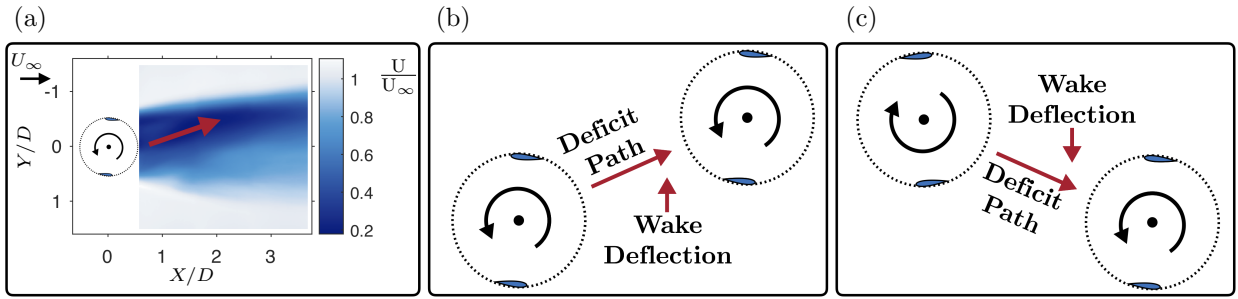

\subsection{Tip-speed ratio control vs. coordinated control}

In \Fref{controlcomp}, the two control strategies are compared for co-rotating and counter-rotating arrays. For most array configurations, the differences between the strategies are limited and without obvious spatial trends. However, regardless of rotation direction for the upstream turbine, when the turbines are aligned in the streamwise direction ($Y/D = 0$), tip-speed ratio control significantly outperforms coordinated control. Since the {downstream turbine operates directly in the wake of the upstream turbine}, tip-speed ratio control allows the downstream turbine to operate at a rotation rate with a more favorable \textit{local} tip-speed ratio (i.e., $\lambda$ defined by the local inflow instead of the freestream flow). Conversely in coordinated control, the turbines are operating at equal rotation rate, which means that the local $\lambda$ for both turbines is far from optimal. 

\begin{figure}
\centering
\vspace{0.2in}
\begin{overpic}[width=\textwidth]{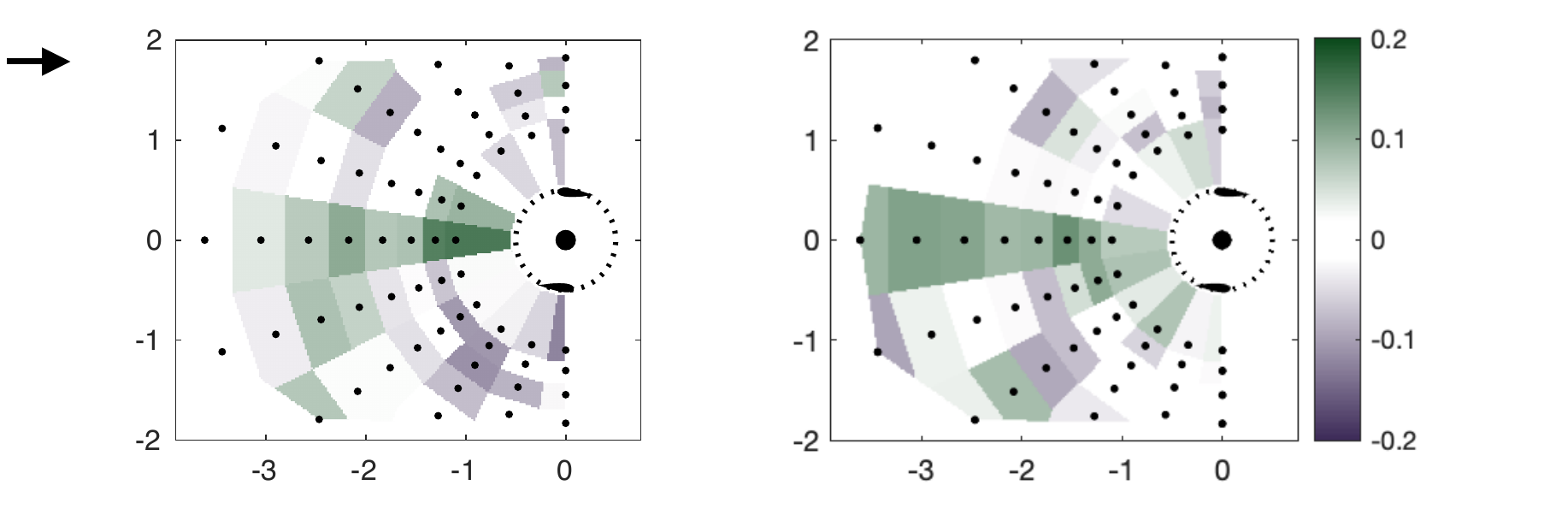}
\put(6,15){\begin{sideways}{$Y/D$}\end{sideways}}
%\put(24.75, -1){{X/D}}
%\put(67.25, -1){{X/D}}
\put(24.75, -0.5){{$X/D$}}
\put(66.5, -0.5){{$X/D$}}
\put(12, 31){Counter-rotating $(\text{TSRC} - \text{CC})$}
\put(58, 31){Co-rotating $(\text{TSRC} - \text{CC})$}
\put(91, 16.5){$ \Delta C_{P,N}$}
\put(5.5, 31){(a)}
\put(48, 31){(b)}
\put(-0.5, 30){$U_\infty$}
\end{overpic}
\caption{Array efficiency comparison ($\Delta C_{P,N} = C_{P,N}^{\text{TSRC}} - C_{P,N}^{\text{CC}}$) of tip-speed ratio control (TSRC) minus coordinated control (CC) for (a) counter-rotating and (b) co-rotating arrays.}\label{fig:controlcomp}
\end{figure}

\subsection{Role of blockage}

\begin{figure}
\centering
\vspace{0.2in}
\begin{overpic}[width=\textwidth]{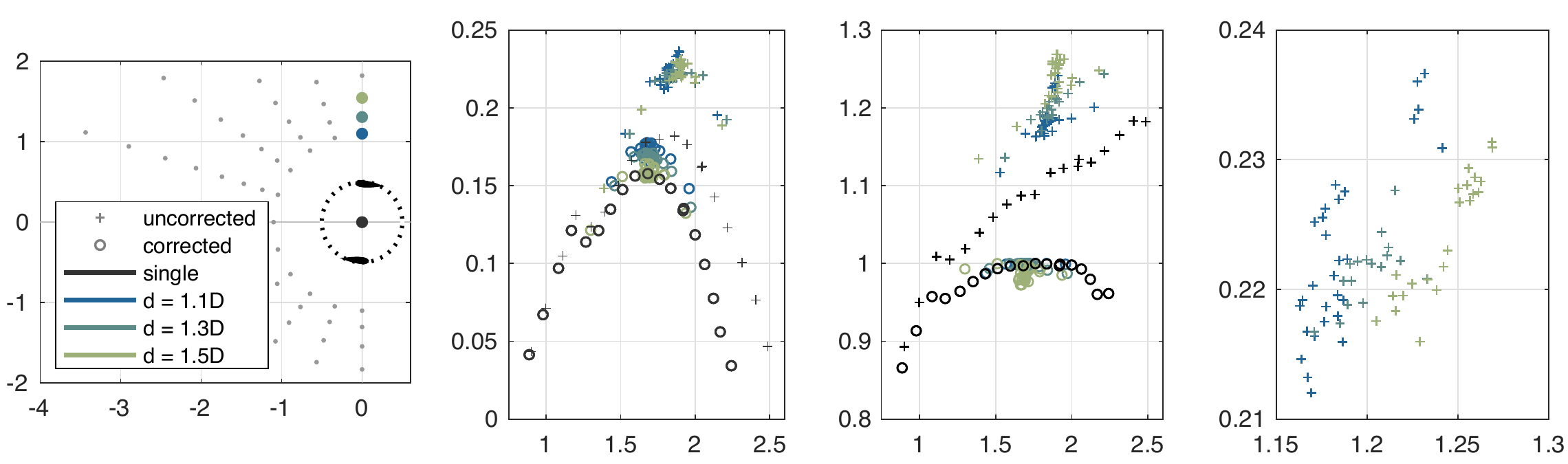}
\put(11.5, 0){{$X/D$}}
\put(-2, 13.25){\begin{sideways}{$Y/D$}\end{sideways}}
\put(46, -0.5){$\ \lambda$}
\put(27.5, 19.5){\begin{sideways}$C_P$\end{sideways}}
\put(70, -0.5){$\lambda$}
\put(52, 19.5){\begin{sideways}$C_T$\end{sideways}}
\put(95, -0.5){$C_T$}
\put(76.5, 21){\begin{sideways}$C_P$\end{sideways}}
\put(3,30){(a)}
\put(33.5,30){(b)}
\put(58.5,30){(c)}
\put(81,30){(d)}
\end{overpic}
\caption{Coefficient of performance(b) and coefficient of thrust (c) for the downstream turbine in three side-by-side array configurations (a) for coordinated control optimization. The point clouds correspond to variations in $\lambda$ and $\phi$ during the control optimization. Values are corrected according to the method of Barnsley and Wellicome~\cite{barnsley1990final}.}\label{fig:blockage}
\vspace{-0.1in}
\end{figure}
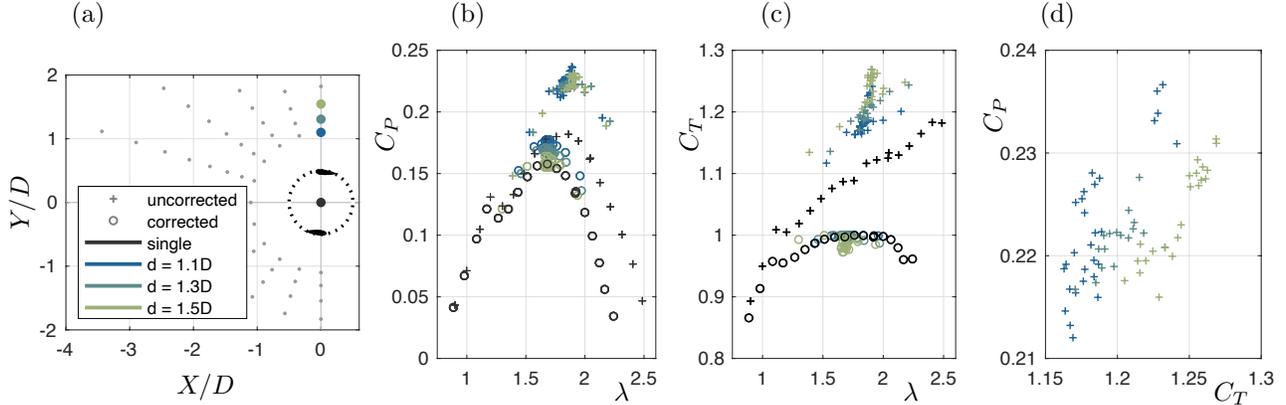

Array configurations have blockages that vary between 2.8\% (turbines in-line) and 5.6\% (turbines side-by-side). {Blockage is well-understood to increase $C_P$ relative to unconfined flow and, using linear momentum actuator disc theory, this can be related to the geometric blockage and thrust coefficient~\cite{ross2020re}. The thrust coefficient, $C_T$, is defined as
	\begin{equation}
	C_T = \frac{T}{\frac{1}{2}\rho U_{\infty}^3 H D}
	\end{equation}
where $T$ is the thrust in the along-channel direction. 

We explore the influence of blockage on our results for the specific case of coordinated control with both turbines at $X/D = 0$, such that the geometric blockage is constant at 5.6\%, but the thrust coefficient changes with $\phi$ and $\lambda$.}
 {Since thrust measurements are only available for the the downstream (stationary) turbine, we focus on its performance at three $Y/D$ locations, denoted by the colored dots in \Fref{blockage}a.}
As shown in \Fref{blockage}b, as the separation distance between the turbines \textit{increases}, the maximum $C_P$ for the stationary turbine \textit{decreases}, which suggests the loss of a beneficial mean or periodic interaction between the turbines. However, as shown in \Fref{blockage}c, $C_T$ also varies substantially during the optimization, likely as a consequence of bluff-body interactions, given variations in apparent solidity for different phase offsets, and $C_P$ is closely correlated with $C_T$ (\Fref{blockage}d). When we correct for this to evaluate the performance in unconfined flow using the method of Barnsley and Wellicome~\cite{barnsley1990final} {(using the blockage for the array and assuming that $C_T$ is similar for the two turbines)}, we see that the corrected thrust coefficients are estimated to be nearly equal to the isolated turbine (colored circles compared to black circles in \Fref{blockage}c). This suggests that, despite the relatively low absolute value of geometric blockage in the experiments, blockage effects may explain much of the observed variation in the thrust coefficient with $\phi$ and $\lambda$ during the optimization process.  
%It is important to note that when blockage is corrected for the stationary turbine in the array that the full array blockage was used to ensure that we were not under-correcting, {\color{red}However, this correction is designed for a single turbine in a channel. It assumes that the thrust coefficient augmentation is due exclusively to blockage. It cannot differentiate between the thrust increase that is due to blockage and a thrust increase that could be due to another turbine in close proximity. This may cause an overcorrection.} 
%I don't agree with the point about thrust increase due to proximity. By varying phi and lambda we are varying the thrust on the array. Blockage corrections are an adjustment for that thrust, regardless of its cause (e.g., turbine control, turbine geometry). 

{However, this does not entirely explain the elevation in $C_P$. Even after accounting for blockage, corrected $C_P$ for the stationary turbine in an array still exceeds corrected $C_P$ for the same turbine in isolation (\fref{blockage}b). This suggests two things. First, the interactions of the two turbines can be optimized to increase their $C_T$, which augments the performance gains from blockage that would occur without coordination. This is an effect that can be meaningfully exploited for cross-flow turbines operating in water. Second, even after we have corrected for blockage, performance enhancements persist and these increase with turbine proximity. This suggests additional mean flow or periodic interaction mechanisms that elevate $C_{P,N}$.} 
	
	 %On the whole, the results presented add to the growing notion that cross-flow turbine arrays can effectively extract energy from wind and marine currents. 
	 %This is kind of a weird place for this sentence - seems more like a concluding thought? Were you trying to get at the contrast with axial-flow turbines?

Finally, we note that, for these tests, the downstream (stationary) turbine is located at the center of the flume, such that the upstream (mobile) turbine has variable proximity to the flume wall depending on configuration. {While this means that wall proximity was not held constant}, given that the flume wall is located at $Y/D \approx \pm 6$, variations in the boundary proximity are unlikely to affect turbine performance~\cite{gauvin2020jrse}. 
\vspace{-0.05in}
\subsection{Experimental design}\label{expdesign}
{The symmetry of the polar grid about the $Y/D$ axis suggests that same parameter space could be sampled more efficiently if only the $Y/D \geq 0$ region is probed.}
%It is tempting to look at the polar grid of geometric configurations in \Fref{paramspace} and deduce that since the locations are symmetric about the $Y/D$ axis and that the same parameter space would be sampled if only the $Y/D \geq 0$ region is probed. 
However, as observed here and for other cases, two arrays with equal $X/D$ spacing and equal and opposite $Y/D$ spacing do not yield identical performance, as a consequence of the asymmetric nature of a cross-flow turbine wake. 

The co-rotating arrays are symmetric when the mobile turbine is at $X/D = 0$  and the $Y/D$ values are equal and opposite (or the absolute center-to-center distances are equal).
Nevertheless, those points are repeated because the two turbines that constitute the array are not identical as discussed in Sec.~\ref{methods}. For side-by-side counter-rotating turbines, no such symmetry exists. In this case, the two possible configurations are characterized by the direction the blades travel as they approach the space between the two turbines.  In one orientation they are traveling upstream, in the other they are traveling downstream, as they approach the space between the two rotors~\cite{zanforlin2016re,vergaerde2018wind,vergaerde2020re}. 

\subsection{Larger arrays}
Depending on the application, arrays of more than two turbines may be desirable. While these experiments consider an array of two cross-flow turbines, the results can be extended to a larger numbers of rotors. The experimental approach presented here would still be possible for a larger parameter space (i.e., each additional turbine adds either a tip-speed ratio or phase to the control scheme). An optimization scheme similar to what is described in this work would be critically important as the number of control variables increases. Additionally, if the free-stream flow speed and direction are uniform across an array at $X/D=0$, the kinematics will be similar for each turbine. Since performance differences between the two strategies are minimal, coordinated control may be a promising option for array design because multiple rotors could be mechanically coupled to a single generator, reducing cost. However, if the array spans heterogeneous (e.g. sheared) inflow, tip-speed ratio control would likely be preferred and could be achieved without an inflow measurement using a non-linear control strategy~\cite{forbush_2019}.

\section{Conclusions}

Experimental control and configuration optimization are performed to maximize array power for a pair of cross-flow turbines. The best-performing arrays have a power coefficient 1.3 times greater than for the turbines operating in isolation. %The hydrodynamic mechanism for the power increase is likely a combination of faster bypass flow into neighboring rotors and {\color{red}proper phase alignment of the rotors}. %since TSR control doesn't control phase, this seems like an over-reach. 
Performance is augmented by three main mechanisms: mean flow alteration {(e.g., faster bypass flow harnessed by the neighboring rotor)}, periodic flow alteration, and blockage. Sometimes they are present individually, but often two or more factors simultaneously affect performance.

This work inspires future studies in several directions. First, there has been some visualization of the interaction mechanisms between cross-flow turbines in dense arrays~\cite{brownstein2019energies}, but they have not been fully visualized and compared to measured performance. {In conjunction with visualization, modeling using recent advances in data-driven methods~\cite{taira2017aiaa,rockwood2017aiaa,noack2011springer} would advance understanding of this system.} {Second, while our results can be provisionally extended to larger arrays, additional interaction mechanisms may occur} when the number of rotors increases beyond two. Third, in this study we present an online experimental optimization for the control strategy. This approach could be applied to other fluid-structure interactions and flow control scenarios~\cite{brunton2015amr} to explore broad parameter spaces in a time-efficient manner. 

\section*{Acknowledgements}

We gratefully acknowledge the support of the Naval Facilities Engineering Command (N00024-10-D-6318 Task Order 0067 and N00024-10-D-6318 Task Order N00024-18-F-8702) and the Army Research Office (ARO W911NF-19-1-0045). The authors acknowledge Eric Clelland and the Bamfield Marine Science Centre for hosting experiments. The authors also acknowledge the support of their colleagues Noah Johnson and Hannah Ross.

\section*{Data availability}

The data that support the findings of this study are available from the corresponding author upon reasonable request.

\begin{appendix}
\setcounter{figure}{0} \renewcommand{\thefigure}{A.\arabic{figure}}
\section{Appendix: Optimized parameters}\label{app}
The optimized parameters for the counter-rotating array operating under coordinated control and tip-speed ratio control are shown in \Fref{co_cc} and \Fref{counter_tsr}, respectively. These results are similar to the co-rotating coordinated control and tip-speed ratio control cases explored in Section \ref{coarrays}. 

%\section{}\label{countertsr}
\begin{figure}[H]
\centering
\begin{overpic}[width=\textwidth]{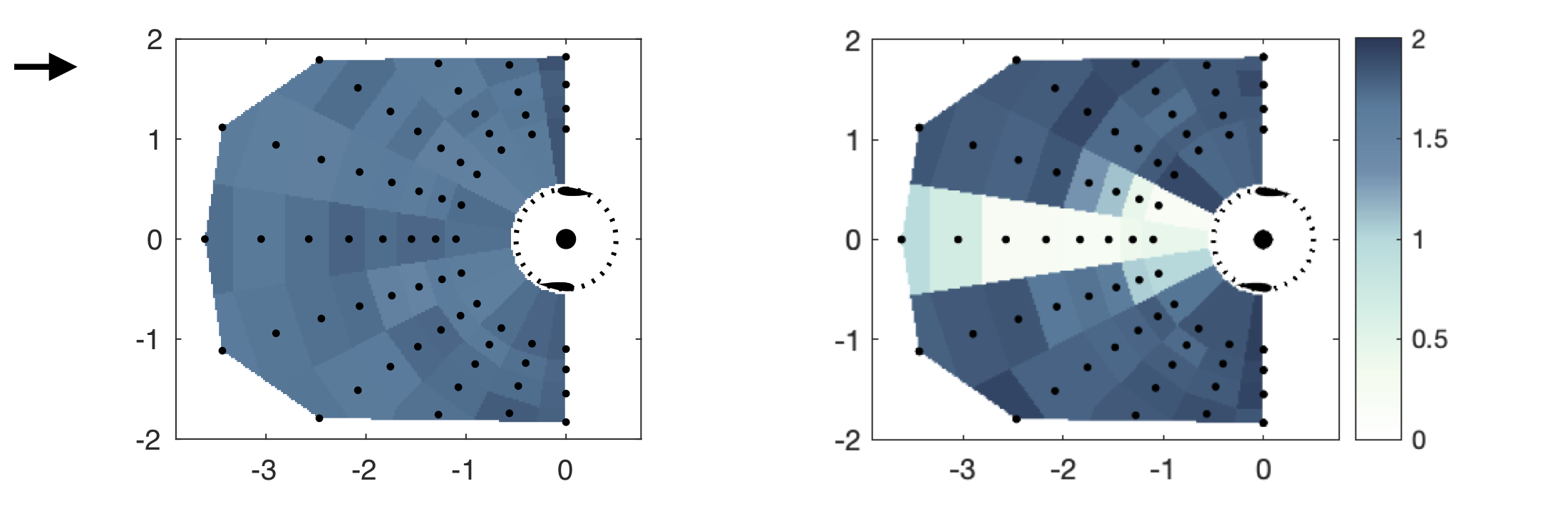}
\put(6,14.5){\begin{sideways}{$Y/D$}\end{sideways}}
\put(24.25, -1){{$X/D$}}
\put(69, -1){{$X/D$}}
\put(5, 30){(a)}
\put(49, 30){(b)}
\put(94, 16.25){$ \lambda$}
\put(-0.5, 30){$U_\infty$}

\end{overpic}
\caption{ Optimized  parameters  for  counter-rotating  turbines  under  tip-speed  ratio  control.   (a)  Upstream turbine tip-speed ratio for a given array layout and (b) downstream turbine tip-speed ratio for the same.}\label{fig:counter_tsr}
\end{figure}

\begin{figure}[H]
\centering
\begin{overpic}[width=\textwidth]{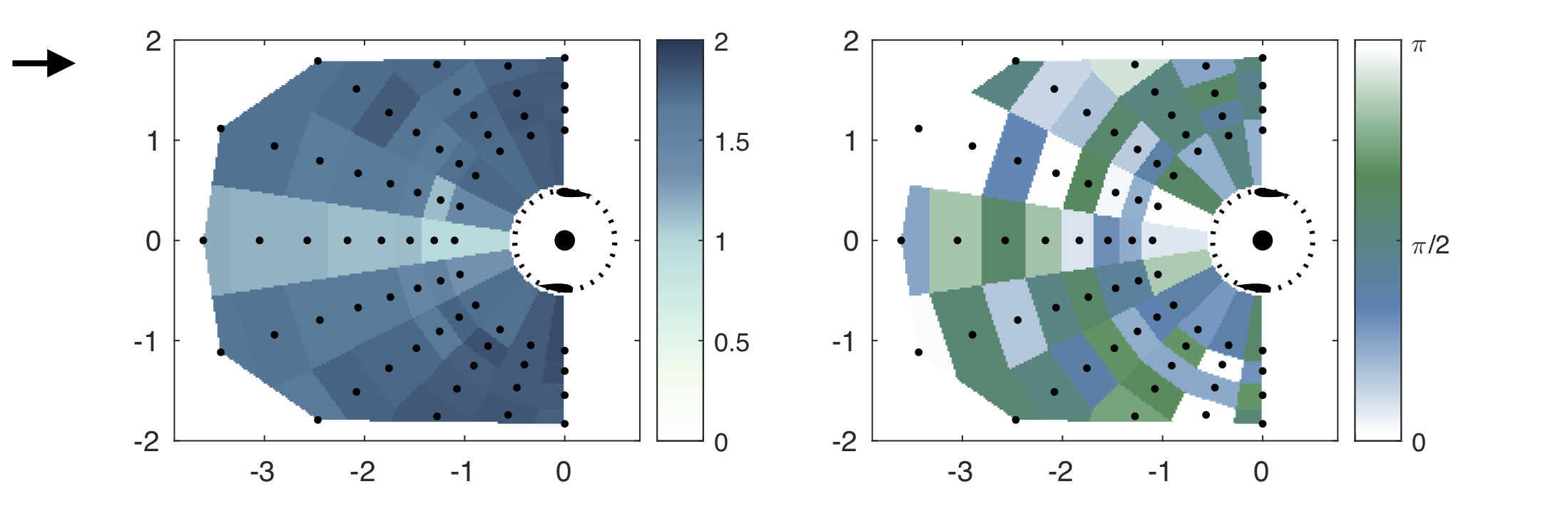}
\put(5,14.75){\begin{sideways}{$Y/D$}\end{sideways}}
\put(24.25, -1){{$X/D$}}
\put(68.75, -1){{$X/D$}}
\put(49.5, 16.25){$ \lambda$}
\put(94, 16.25){$ \phi$}
\put(5.25, 29.5){(a)}
\put(50, 29.5){(b)}
\put(-0.5, 30){$U_\infty$}
\end{overpic}
\caption{Optimized parameters for counter-rotating turbines under coordinated control: (a) optimized tip-speed ratio for the turbine pair and (b) optimized phase difference.}\label{fig:counter_cc}
\end{figure}

\end{appendix}

\bibliographystyle{plain}
\bibliography{denseArrays}
\end{document}